 \documentclass[authoryear,preprint,review,12pt]{elsarticle}

\usepackage{amssymb}
\usepackage{graphicx}

\def\astrobj#1{#1}
\journal{New Astronomy}

\begin{document}

\begin{frontmatter}



\title{New Photometric Study of A Southern Contact Binary System: \astrobj{HI~Puppis}}


\author[itk]{B. Ula{\c s}\corref{dip}}
\author[saa,gau]{{C. Ulusoy}}
\address[itk]{\.{I}zmir Turk College Planetarium, 8019/21 sok., No: 22, \.{I}zmir, Turkey}
\address[saa]{College of Graduate Studies, University of South Africa, PO Box 392, Unisa, 0003 Pretoria, South Africa}
\address[gau]{Girne American University, Aviation Management, PO Box 388, Girne, North Cyprus}
\cortext[dip]{Corresponding author \\
E-mail address: bulash@gmail.com}

\begin{abstract}
We present results of new photometric observations of the contact binary system \astrobj{HI~Pup} as well as the radial velocity curve of the system. Time series multicolour photometry was obtained at the South African Astronomical Observatory (SAAO) using the 1-m Cassegrain Telescope. We applied a simultaneous solution to the $BVRI$ light and radial velocity curves in order to determine the physical parameters of the system. From an analysis of the new multicolour data, the physical parameters were found to be $M_1=1.22M_{\odot}$, $M_2=0.23M_{\odot}$, $R_1=1.44R_{\odot}$, $R_2=0.67R_{\odot}$, $L_1=3.3L_{\odot}$, $L_2=0.7L_{\odot}$. Our solution confirms that HI Pup has a typical A--type W~UMa binary system characteristics.
\end{abstract}


\end{frontmatter}

\section{Introduction}

The variability of \astrobj{HI~Pup} (=HIP~36762) was first discovered by \cite {hof49}. \cite{sah63} classified this target as an eclipsing binary sytem showing a W|UMa type light curve. Furthermore, they reported \astrobj{HI~Pup} to be a possible member of galactic cluster \astrobj{Cr~173}, with a period of 0$^d$.4. The first photoelectric light curve of HI Pup  was obtained by \cite{ker84}. They confirmed W~UMa--type variation in the light curve of the system  and found the orbital period to be 0$^d$.43257. The star was in the field of {\it HIPPARCOS} satellite \citep{esa97} \cite{ruc97} calculated absolute magnitude of the system using the data taken by {\it HIPPARCOS}. \cite{ade01} listed the {\it HIPPARCOS} targets that have largest photometric amplitudes. \cite{pri03} included \astrobj{HI~Pup} into their catalogue of the field contact binary stars. \cite{due07} obtained and fitted the radial velocity curve of the star. The spectroscopic mass ratio of the system was calculated to be 0.19 by the authors. The physical parameters of 62 binaries including \astrobj{HI~Pup} were calculated by \cite{deb11} using the ASAS (All Sky Automated Survey, \cite{poj03}) database. They obtained the light elements and the absolute parameters of the system by using the radial velocity curve of \cite{due07}.

In this paper, we present the details of our observations and reduction process in the next section. The simultaneous light and radial velocity curve analyses are explained in Sec. 3. We briefly discuss our results in the last section.

\section{New Observations and Data Reduction}

Multicolour CCD photometry was used to obtain the light curve and light elements of the system. Observations were carried out with the 1-m Elizabeth Cassegrain telescope at the South African Astronomical Observatory (SAAO) in Sutherland station, South Africa. The new multicolour data were acquired with the instrument STE-4, a 1024x1024 pixel back illuminated CCD, through Bessel type $BVRI$ filters in January 2014 (Julian dates between 2456674.3405 and 2456679.5196). During the observations, a total of 1518 observing points in the $B$, $V$, $R$ and $I$ filters were collected.

Standard {\tt IRAF} steps including correction for dark, bias and flat-fielding of each CCD image were followed for data reduction. In order to measure instrumental magnitudes for stars in the CCD field aperture photometry technique was applied by using the {\tt DAOPHOT II} software package \citep{ste87}.We used \astrobj{TYC~8141-1388-1} and \astrobj{TYC~8141-1374-1} as comparison and check stars, respectively. The observed light curves of the system in the $BVRI$ filters can be seen in Fig.~\ref{figlc}. Three times of minimum were also computed by using Kwee--van Woerden method \citep{kwe56} and listed in Table~\ref{tabmin}.

\begin{table}
\caption{Derived minimum times calculated from observations.}
\label{tabmin}
\begin{tabular}{ll}
\hline
\multicolumn{1}{c}{HJD} & Type      \\
\hline
2456677.4087(1)	    	  & \multicolumn{1}{c}{II} \\
2456678.4883(3)           & \multicolumn{1}{c}{I} \\
2456679.3530(3)           & \multicolumn{1}{c}{I}\\
\hline
\end{tabular}
\end{table}

\section{Solution of Light and Radial Velocity Curve}

The first light curve solution of the system was obtained by \cite{ker84}. The authors analyzed the light curve by means of Wilson--Devinney \citep{wil71} and Wood \citep{woo72} methods. They calculated the difference between the two minimum  0$^m$.2. \cite{due07} obtained the radial velocity curve of the system and   derived the orbital parameters $V_{0}$=59.9~km/s, $K_1$=50.2~km/s, $K_2$=265.2~km/s. \cite{deb11} analyzed the  ASAS-3 light curves of \astrobj{HI~Pup} aimed at determination of the physical parameters. \cite{deb11} improved the orbital period given in the ASAS database by following the minimisation of entropy method.

\begin{figure}
\includegraphics{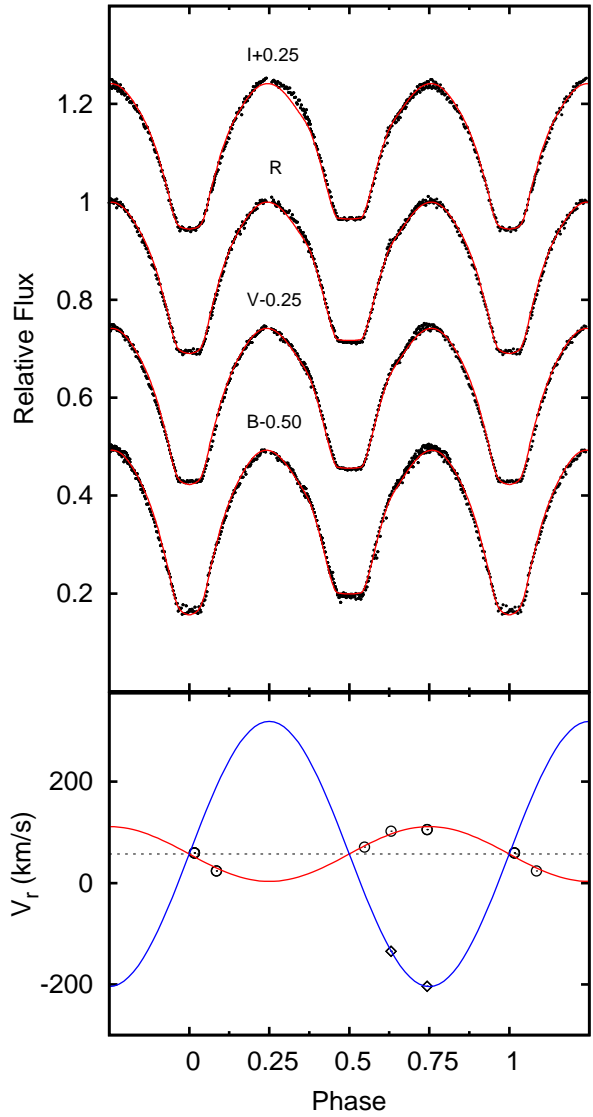}
\caption{The agreement between observations and the result of the analysis for light (top) and radial velocity (bottom) curves. The radial velocity data are from \cite{due07}.}
\label{figlc}
\end{figure}

\begin{figure}
\includegraphics{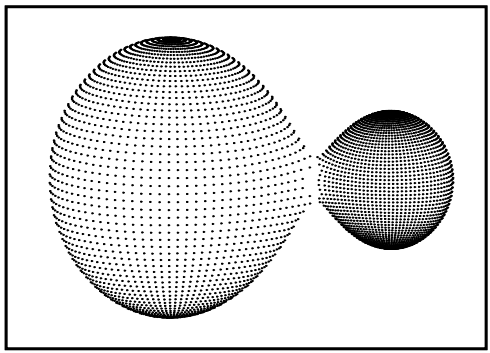}
\caption{The geometric configuration of the system as seen at the phase $\phi$=0.25.}
\label{figgeo}
\end{figure}

For our study, we analyzed our $BVRI$ light curves and combined with the radial velocity data of \cite{due07} simultaneously. In order to perform our analysis, the {\tt PHOEBE} \citep{prs05} software , which is based on the Wilson--Devinney code \citep{wil71}, was used. The software was run in the suitable mode for contact binaries with different temperatures. The inclination $i$, temperature of secondary component $T_2$, surface potential $\Omega_1 = \Omega_2$, mass ratio $q$, velocity of the centre of mass $V_0$, semi--major axis $a$, luminosity of the primary component $L_1$, the time of minimum $T_0$ and the orbital period $P$ were set as free parameters during the solution. The initial value of the mass ratio was adopted by following \cite{due07}. The temperature of the primary component was fixed to 6500 K using the correlation given by \cite{cox00} due to its spectral type (F6V, \cite{due07}). Our resulting parameters are listed in Table~\ref{tablc}. The theoretical light and radial velocity curves are illustrated in Fig.~\ref{figlc}. The geometric configuration of the system is also plotted in Fig.~\ref{figgeo}.

The light curves of HI Pup obtained from the current data slightly show magnitude difference between two maximum in the $B$ and $V$ filters. This phenomena can occur if surfaces of the components are covered by stellar spots. Because of this reason, we applied another simultaneous analysis with the assumption of a cool spotted region on the cooler component. Nevertheless, the deviation between the observations and theoretical light curve increased, particularly in the $R$ and $I$ filters, as the fixed spot parameters were adjusted. The program therefore did not converge any reasonable result that can be attributed to a stellar spot. Yet, the $B$ light curve individually analyzed by setting at fixed cool spot on the secondary component. The smallest standard deviation was reached at the following spot parameters; co-latitude $\beta$=40$^o$, longitude $\lambda$=90$^o$, radius $r$=20 and temperature factor $t$=0.9.

In conclusion, we find that the results given in the previous study by \cite{deb11} differ from our results concerning the mass ratio and the degree of contact factor. Morever, \cite{deb11} derived the temperatures as $T_1$=6514~K and $T2$=6662~K for primary and secondary components, respectively and our results show that the primary component is hotter than the secondary one. For the determination of the spectral type of the secondary component we assumed that HI Pup is located on the main sequence in the HR Diagram and the component is an F7 star regarding as the correlation relation given by \cite{cox00}.

\begin{table}
\caption{Result of the light curve analysis. Formal 1$\sigma$ errors are given in paranthesis.}
\label{tablc}
\begin{tabular}{ll}
\hline
Parameter                                   & Value   \\
\hline
$i$ ${({^\circ})}$                          & 82.2(9)  \\
$q$                                         & 0.206(1)  \\
$V_{0}$~(km/s)				    & 57.3(2)	\\
$a$ (R$_\odot$)				    & 2.72(1) \\
$\Omega _{1}=\Omega _{2} $                  & 2.221(32)    \\
$T_1$~(K)                               & 6500   \\
$T_2$~(K)                               & 6377(24) \\
Fractional radius of primary                & 0.528(10)\\
Fractional radius of secondary              & 0.262(12) \\
Luminosity ratio:$\frac{L_1}{L_1 +L_2}$ &   \\
$B$                                         & 0.820(58)        \\
$V$                                         & 0.816(49)        \\
$R$                                         & 0.814(42)         \\
$I$                                         & 0.812(37)       \\
\hline
\end{tabular}
\end{table}

\section{Results and Conclusion}

\begin{figure}
\includegraphics{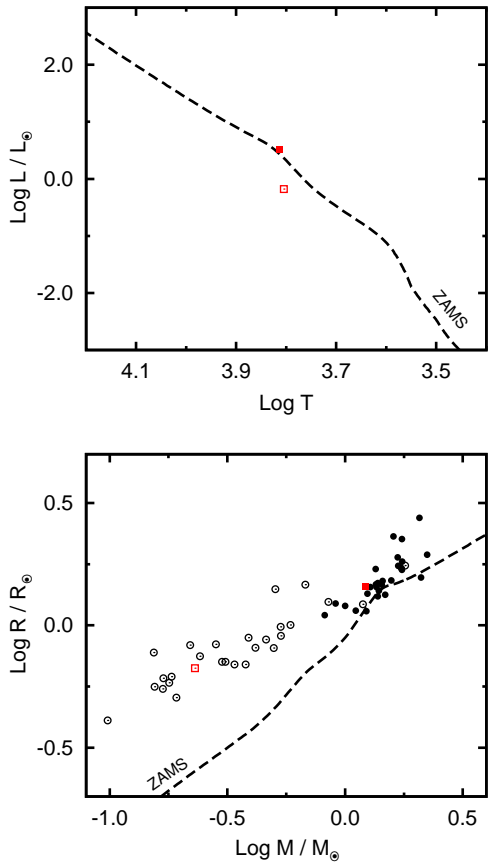}
\caption{Location of the components on the Hertzsprung--Russell diagram (top). The components are placed on the mass--radius plane with various A--type W~UMa binaries given by \cite{lie08} at the bottom panel. Primary components are symbolised by filled circles while the secondaries are represented with open circles. Squares correspond to components of \astrobj{HI~Pup}. ZAMS data are taken from \cite{pol95}.}
\label{fighr}
\end{figure}

We analyzed the new multicolour light and radial velocity \citep{due07} curves of \astrobj{HI~Pup} to derive the physical parameters of the system that are listed in Table~\ref{tababs}. The location of the components, corresponding to the determined physical parameters, in the HR diagram is shown in the top panel of Fig.~\ref{fighr} The less massive and cooler component is located below the Zero Age Main Sequence (ZAMS) while the primary component is close to the secondary one.

The degree of contact of the system, $f$, was calculated by following the equation $f=\frac{\Omega_{I}-\Omega}{\Omega_{I}-\Omega_{O}}$ where $\Omega_{I}$ and $\Omega_{O}$ are inner and outer Lagrangian surface potential values, respectively. The result is found to be $f$=0.2.

\cite{bin70} reported that the occultation of the massive component during the primary minimum is a characteristic property of the light curves of A--type W~UMa. The author also categorized two sub-classes of W~UMa systems based on the spectral types of the systems: (i) A--type W~UMa systems which lay between A9 and F8 and (ii) W--types which appear in the spectral type range F7--M5. The evolutionary status of these two type of W~UMa systems are remarkably different from each other. Early studies propose that A--type W~UMa systems are more evolved than W--types. However, \cite{mac96}, \cite{gaz06} proved that the idea was no longer accepted. \cite{mac96} suggested three class for contact binaries: young late--type contacts, old late--type contacts and the rest of contact systems. They remarked that the young and old late--type contact binaries can be considered as W~UMa type. However, third class are related to hot contact systems and they have different evolutionary status. \cite{gaz08} suggested that these two sub-types should probably belong the group of cool contact binaries and  A--type W~UMa system with high mass can preserve its contact configuration during its evolution. Medium mass W--type binaries, on the other hand, evolve to A--type systems with extreme mass ratios.

The massive component is occulted at primary minimum in the case of HI~Pup. Therefore, the system shows a typical A--type W~UMa light curve. The spectral type given by \cite{due07}, F6, also prove its membership of A--type according to the classification of \cite{bin70}. We marked the components of \astrobj{HI~Pup} with the components of other A--type systems \citep{lie08} on the mass-radius plane. In the bottom panel of Fig.~\ref{fighr} it can be clearly seen that the primary and secondary components of A--type W~UMa systems gather in two separate locations in mass--radius diagram.

\begin{table}
\caption{The derived absolute parameters of the system with the analysis. The effective temperature of the sun is adopted to 5780~K during the calculations.}
\label{tababs}
\begin{tabular}{lrr}
\hline
Parameter                &\multicolumn{2}{c}{Value}    \\
& \multicolumn{1}{c}{Pri.}&\multicolumn{1}{c}{Sec.} \\
\hline
M (M$_{\odot}$)          & 1.21(24)         & 0.23(19) \\
R (R$_{\odot}$)          & 1.44(11)          & 0.67(10)\\
L (L$_{\odot}$)		 & 3.3(5)	     & 0.7(2) \\
M$_{bol}$ ($^m$)         & 3.4(9)	     & 5.2(1.7) \\
$a$ (R$_{\odot}$)        & \multicolumn{2}{c}{2.7(2)}  \\
\hline
\end{tabular}
\end{table}

\section*{Acknowledgements}
BU thanks to scholar group A.S.E.K. CU wishes to thank the South African National Research Foundation NRF MULTI-WAVELENGTH ASTRONOMY RESEARCH PROGRAMME (MWGR), Grant No: 87635. This work is based upon research supported by the National Research Foundation and Department of Science and Technology of the Republic of South Africa. This paper uses observations made at the South African Astronomical Observatory (SAAO). This study made use of IRAF Data Reduction and Analysis System.

\end{document}